\documentclass[
  reprint,
  amsmath,amssymb,
  aps,
  showkeys
]{revtex4-2}

\usepackage{graphicx}
\usepackage{dcolumn}
\usepackage{bm}
\usepackage{comment}
\usepackage{float} 
\usepackage{lipsum}
\usepackage[caption=false]{subfig}
\usepackage{booktabs}
\usepackage[utf8]{inputenc}
\usepackage[T1]{fontenc}
\usepackage{mathptmx}
\usepackage{parskip} 
\usepackage{microtype}

\begin{document}

\preprint{KR_PRN/123-QED}

\title{Prospects for market-specific design of Perovskite-Silicon tandem solar cells}

\author{Karthik Raitani}
\email{karthikr@iitb.ac.in}

\affiliation{Department of Electrical Engineering, Indian Institute of Technology Bombay, Mumbai 400076, Maharashtra, India}

\author{Pradeep R. Nair}
\email{prnair@ee.iitb.ac.in}

\affiliation{Department of Electrical Engineering, Indian Institute of Technology Bombay, Mumbai 400076, Maharashtra, India}
	
	\begin{abstract} 
		The quest for optimal perovskite for tandem cell configurations is challenging as it involves several factors ranging from device level performance under field conditions to degradation rates and cost. Here, we first highlight the limitations of traditional detailed balance or Shockley-Queisser (SQ) analysis towards the design of Perovskite/Silicon tandem solar cells. Through well-calibrated numerical simulations, we evaluate geographic location-specific annual energy yield (EY) and quantify the influence of temperature-dependent material and transport parameters. Our results indicate that the EY scales in a near-identical manner with the top cell band gap ($E_{gT}$) for various geographic locations. In comparison to SQ analysis, our simulations predict a twofold relaxation in the target degradation rates at which perovskites over a broad range of band gaps could yield a comparable levelized cost of Energy (LCOE). These insights are of broad interest for the development of perovskite materials, and test protocols to evaluate the stability of Perovskite-Silicon tandem solar cells.  
     \end{abstract}

	\maketitle
	
	\section{Introduction}

	Tandem photovoltaics, with its promise of high efficiency and cost competitiveness, offers new possibilities to advance the solar industry \cite{de2023tandems,alberi2024roadmap}.	Among the various competing technologies, Perovskite-Silicon (P-Si) two terminal (2-T) tandem solar cells (TSCs) have emerged as a front-runner with  efficiencies beyond the classical single-junction theoretical limit \cite{liu2024perovskite,pathways}. Along with better absorption properties, perovskites offer bandgap ($E_g$) tunabilty over a broad range \cite{liu2023bimolecularly,yang2018high,ramadan2023methylammonium}.   
     For tandem applications, it is well known that $1.6-1.8\, \mathrm{eV}$ is the suitable $E_{g}$ range for  perovskites \cite{unger2017roadmap,wang2023suppressed}.    Metal halide perovskites with $E_{g}$ above $1.6 \, \mathrm{eV}$ typically require a higher bromine content ($\mathrm{Br}\,,\%$), which could induce halide phase segregation and associated stability/degradation issues \cite{caprioglio2023open, yao2025oriented}. 
	As such, identification of the optimal top cell material for P-Si tandem solar cells which yields a competitive LCOE turns out to be a very complex optimization problem \cite{chang2021bottom,ahangharnejhad2022impact,sofia2020roadmap,yu2018techno}. Among the various factors, LCOE depends on the degradation rate, lifetime, cost, and energy yield. Given that the stability concerns in perovskites are yet to be fully addressed \cite{stability_challenges,boyd2018understanding,fu2022monolithic,jiang2024rapid}, it is imperative that certain benchmark criteria are identified which allows early as well accurate quantitative guidelines for material/process development.\\ 
        	\begin{figure}[h!]
		\centering
		\includegraphics[width=0.93\linewidth]{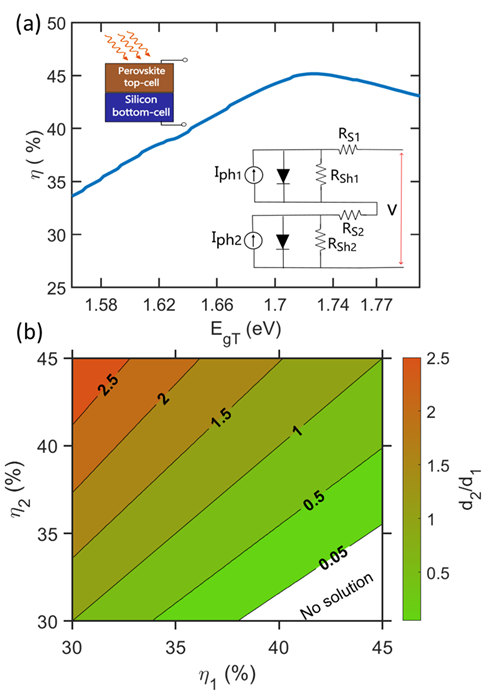}
		\caption{
         Location‑agnostic, idealized phase map illustrating performance–stability trade‑offs in 2-T Perovskite-Silicon tandem solar cells. (a) Efficiency limits for 2-T  tandem cells (at $300\,\mathrm{K}$) for different top cell bandgaps ($E_{gT}$) as per Shockley-Queisser (SQ) analysis. The insets show a schematic of P-Si 2-T tandem solar cell and its equivalent circuit model. (b) Phase map of the ratio of degradation rates ($d_2/d_1$) at which a device with  $\eta_2$ has the same LCOE as a device with $\eta_1$ ($d_1=0.03$, $r=0.065$ and $N=25$ years. Other parameters are assumed to be the same).} 
		\label{fig:phase}
	\end{figure}
    \vspace{1cm}
     Shockley-Queisser (SQ) analysis predicts that the maximum achievable efficiency for 2-T TSCs with a Si-based bottom cell is around 45\% \cite{de1980detailed,ruhle2016tabulated}. This ideal efficiency could be achieved using top cells with $E_{gT} \sim 1.73\, \mathrm{eV}$  (see Fig. \ref{fig:phase}a. Detailed description of the methodology is available in  Section 1, Fig.~S1 of the Supporting Information). The same analysis also indicates that the efficiency of 2-T TSCs is extremely sensitive to the $E_{gT}$ - i.e., the efficiency  decreases by ~7\% per $0.1\, \mathrm{eV}$ decrease in the $E_{gT}$ from its optimal value (see Fig. \ref{fig:phase}a). In relative terms, the efficiency drops by about 15\% of its maximum value as the $E_{gT}$ varies by $0.1\, \mathrm{eV}$.  This extreme sensitivity on the $E_{gT}$ raises several system-level concerns: (a) Is it a foregone conclusion that a perovskite with $E_g \sim 1.73\, \mathrm{eV}$ is the optimal choice for 2-T configurations?  (b) Does the EY of P-Si 2-T TSCs also show a similar sensitivity on the $E_{gT}$? (c) Are there any scenarios in which 2-T TSCs with a sub-optimal $E_{gT}$ can be competitive, especially given the fact that large $E_{g}$ perovskites are more susceptible to stability issues related to light induced phase segregation \cite{caprioglio2023open, yao2025oriented} (d) How does geographic location and climatic conditions impact the above? Or, what would be the scope for climate-specific design of P-Si 2-T TSCs?\\
	
	The choice of the optimal perovskite for 2-T P-Si TSCs depends on the corresponding estimates for LCOE  \cite{de2023levelized} which is defined as 
    \begin{equation}
   LCOE= \frac{\text{Total lifetime cost}}{\text{Total lifetime energy yield}}
    \label{L1}
\end{equation}
For P-Si 2-T tandem solar cells, it is appropriate to assume that the lifetime cost at a given location is independent of the $E_{gT}$. This follows from the fact that the bandgap can be tuned via compositional engineering without significantly impacting manufacturing cost and the balance of system costs are also expected to be independent of $E_{gT}$. Under such conditions, two technologies with same lifetime EY result in similar LCOE. Accordingly, the necessary condition for two different technologies with similar lifetime cost to have the same LCOE is
	
        \begin{align}
    \text{Total lifetime EY} _{1} &= \text{Total lifetime EY} _{2} \nonumber \\
    \sum_{n=1}^{N} EY_{1}(1-d_1)^{n}(1+r)^{-n} &= \sum_{n=1}^{N} EY_{2}(1-d_2)^{n}(1+r)^{-n}
    \label{eq:criteria}
\end{align}
	where $EY$, $d$, and $r$ denote the initial annual energy yield, degradation rate, and discount factor, respectively. The subscripts identify the respective technologies.\\	
    
	For a given $N$ and $r$, eq. \ref{eq:criteria} indicates that two different technologies could yield same LCOE for several combinations of $EY$ and $d$.  Under the assumptions that the temperature coefficients, incident spectrum, and ambient conditions (temperature, wind speed, etc.) are the same for both technologies (i.e., at a given location), we have $EY \propto \eta$, where $\eta$ is the efficiency. Accordingly, the variables of interest are, as per eq. \ref{eq:criteria}, $\eta$ and $d$. Fig. \ref{fig:phase}b shows a phase-plot of different scenarios in which two technologies with different efficiencies could yield the same LCOE as a function of $\eta$ and $d$ (with $N=25$ years, $r=6.5\%$, and $d_1=3\%$ per year. All other parameters are the same). Here, the color plot indicates the ratio $d_2/d_1$ for which eq.~\ref{eq:criteria} is satisfied.\\
    
    Fig. \ref{fig:phase}b indicates that a technology with slightly inferior efficiency could still emerge competitive if it has a lower degradation rate. For example, as per the trends shown in Fig. \ref{fig:phase}b, a new technology with $\eta_2=32\%$ is better than the baseline with $\eta_1=40\%$ if $d_2 < 0.1d_1$.  On the other hand, a new technology with $\eta_2=40\%$ is better than the baseline with $\eta_1=32\%$ for $d_2<2d_1$. Interestingly, there are scenarios in which the LCOEs of two technologies are not comparable. Such cases are shown by the white triangle in Fig.~\ref{fig:phase}b. Using the SQ limit efficiencies (see Fig. \ref{fig:phase}a), these arguments indicate that TSCs with $E_{gT,2}=1.62\, \mathrm{eV}$ can be competitive with TSCs having $E_{gT,1} = 1.71\, \mathrm{eV}$ if $d_2<0.3d_1$. Now, is this a reasonable estimate? What would be the influence of geographic locations and climatic conditions on this selection criteria?
	
    In this manuscript, we address the above-mentioned system-level concerns associated with P–Si 2-T tandem solar cells and identify geographic location and climate-aware guidelines for selecting the optimal perovskite material. Section~II compares SQ limits of 2T P-Si tandem solar cells with numerically simulated efficiencies. Section~III elaborates on the annual energy yield and levelized cost of electricity (LCOE) of 2-T TSCs for various $E_{gT}$; and Section~IV elucidates the influence of degradation rates and location specific criteria for optimal $E_{gT}$. Our numerical simulations account for temperature-dependent optical absorption, carrier transport, and recombination under field conditions. Our results indicate that the estimates based on the SQ analysis tend to be overly conservative. In addition, we find that despite enormous variations in climatic conditions, normalized EY shows similar scaling behavior with $E_{gT}$ for several cities throughout the world. Further, a wide variety of perovskite materials are suitable as top cells in 2-T TSCs with appropriate improvements in degradation rate. There is no single globally optimal material; indeed, the field of material research for optimal perovskites remains open and is expected to continue evolving in the near future.\\

	\section{SQ limits vs. Practical estimates }
	
	Tandem solar cells can be understood in terms of series connection of the top cell and the bottom cell (see inset of Fig. \ref{fig:phase}a). The current density vs. voltage (J-V) characteristics of each of the individual sub-cells, in the absence of injection currents, series $(R_{s})$, and the shunt resistances $(R_{sh})$, can be described as
	
	\begin{equation}
		J=-q\int \alpha\Phi_{in}dE +qR_T
		\label{eq:JV}
	\end{equation}
	
	where the first term on the RHS denotes the current due to the photo-generation and the second term denotes the net recombination of the generated photo-carriers. Here, $\alpha$ is the absorption probability, $\Phi_{in}$ is the incident spectrum, $E$ is the energy, and $R_T$ is the net recombination rate (i.e., per unit area, per unit time). Note that while $\Phi_{in}$ for the top cell is the location specific incident spectrum, the photogeneration in the bottom cell is due to the transmitted spectrum from the top cell. Evidently, being a series connection, the efficiency of 2-T tandem solar cells is critically influenced by the current mismatch between the top and the bottom cells \cite{aydin2020interplay}. However, this current mismatch is not a static aspect and in turn depends on several factors (incident spectra,  temperature dependence of $E_{g}$, optical constants, etc.) \cite{nkconstant}.\\
	
	SQ analysis has certain limitations as it assumes that (i) the spectrum incident on the top cell is $AM1.5G$, and (ii) each sub-cell absorbs all available photons which are more energetic than its $E_{g}$ (i.e., $\alpha=0$ for $E<E_g$, and $\alpha=1$ for $E>E_g$) \cite{sqlim}. In fact, the absorption probability can be less than unity due to the material properties, finite thickness of the active layer, etc. Further, several parameters like refractive indices, $E_{g}$, and recombination coefficients are temperature dependent. In addition, Perovskite and Silicon bandgaps have contrasting temperature dependence - the former increases while the latter decreases with temperature \cite{moot2021temperature}.  Another major limitation is that SQ analysis accounts only for the fundamental radiative recombination in both subcells. However,  practical solar cells are often limited by non-radiative recombination. Further, the efficiency could be constrained by other aspects in a solar cell like transport layers, band offsets, ion migration, etc. \\ 
	
	The above-mentioned limitations can be addressed through numerical simulation of 2-T cells. The methodology of the same is detailed in Section 2 of Supporting Information. Briefly, we rely on self-consistent numerical solution of continuity equations for mobile carriers and ions along with Poisson's equation for electrostatics to arrive at the JV characteristics for the top subcell. The carrier generation rate is calculated using the Transfer Matrix Method (TMM) \cite{transfer} accounting for optical losses with experimentally determined refractive indices \cite{nkconstant} (See Fig.~S2-S4 for more details). We consider trap assisted (mono molecular), radiative (bi-molecular), Auger, and interface recombination in our simulations. \\
	
	The Si bottom cell is represented in terms of an equivalent circuit with calibrated temperature coefficients. The JV characteristic of the tandem cell is then obtained as a series connection of top and bottom cells  (see Fig.~S7 and Section 2.4 of Supporting Information). The temperature dependence of parameters like refractive indices and $E_{g}$ are explicitly taken into account in our simulations.  This modeling framework is well calibrated with experimental results and our previous publications have addressed several aspects like ion migration, large signal switching, and phase segregation in perovskite based solar cells through such self-consistent numerical simulations \cite{sivadas2023ionic,singareddy2021phase,chittiboina2023intrinsic,agarwal2015device,saketh2021ion}.\\
	
\begin{figure}[ht!]
		\centering
		\includegraphics[width=1.0\linewidth]{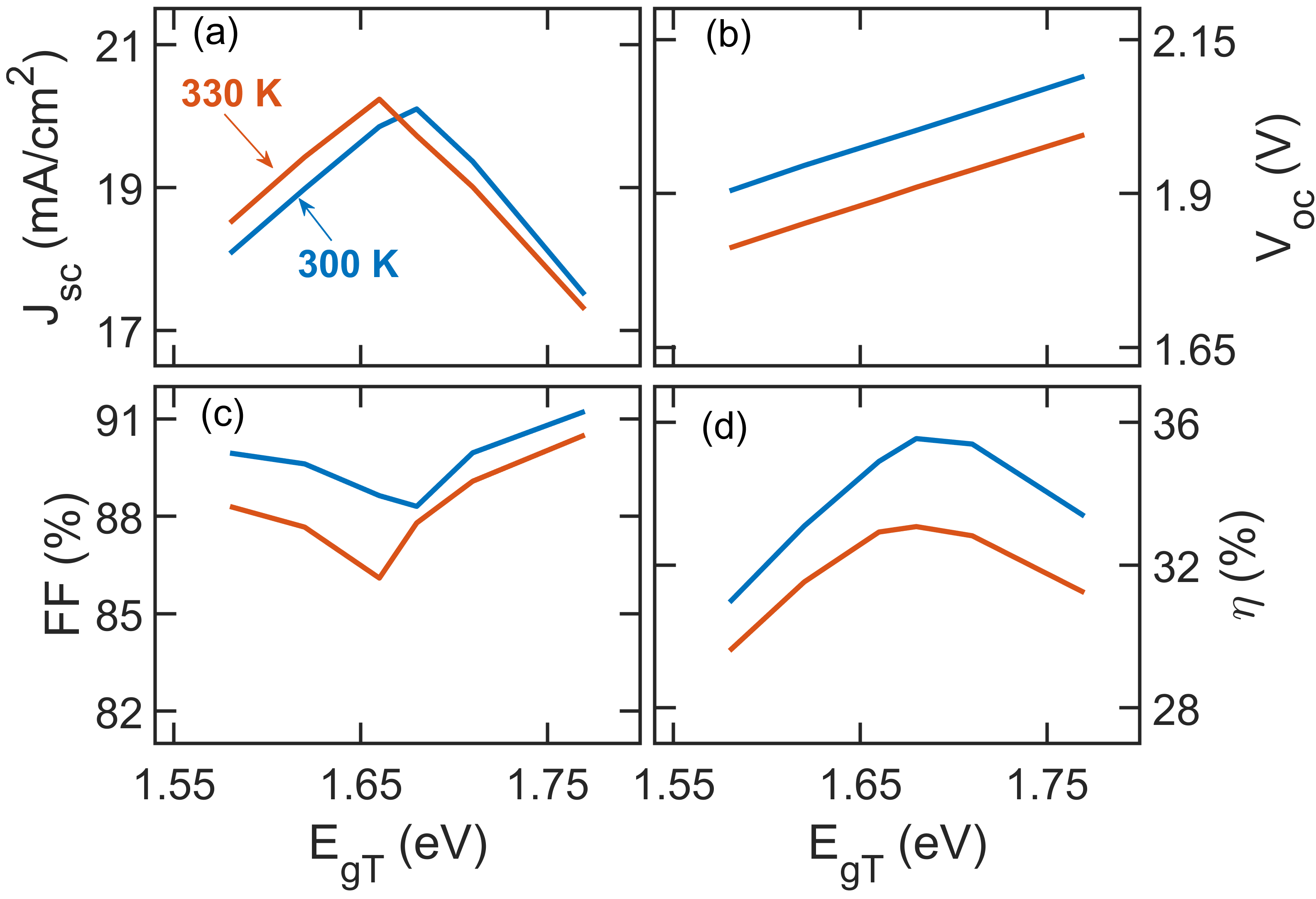}
		\caption{ Numerical simulation results on the temperature dependent parameters of P-Si 2-T tandem solar cell at 300 and 330 K. (a) Variation of $J_{sc}$, (b) $V_{oc}$ and (c) Fill factor (FF) and (d) efficiency ($\eta$) with $E_{gT}$.}
		\label{eff_estimate}
	\end{figure}

	Fig. \ref{eff_estimate} shows the variation of key performance parameters of 2-T TSCs as a function of $E_{gT}$. The set of parameters used in this study (and their temperature dependence) are provided in Table~S2 and S3 of Supporting Information. Our simulations compare well with the experimental results available in literature. For example, Liu \textit{et al.}, reported excellent performance for 2-T P-Si TSCs \cite{liu2024perovskite} with parameters $J_{SC}=20.67\, \mathrm{mA/cm^2}$, $V_{OC}=1.98\,  \mathrm{V} $, and $\eta=34.08\%$. In comparison, our simulations predict $J_{SC}= 20.1\, \mathrm{mA/cm^2}$, $V_{OC}= 2.0\, \mathrm{V} $ and $\eta=35.5\%$ for 2-T TSCs with $E_{gT}=1.68\, \mathrm{eV}$ - thus validating the simulation methodology and parameters used. A comparison of the simulated JV characteristics with experimental results from the above reference is provided in Fig.~S6 and Table~S1 of Supporting Information.  Parasitic series resistance effects are ignored in our numerical simulations which explains the higher efficiency as compared to literature reports. In fact, recent literature  attributes almost $2\%$ absolute loss in $\eta$ to series resistance - which is consistent with this work \cite{ugur2024enhanced}.  Fig.~S9 of Supporting Information explores the influence of recombination parameters on the efficiency of 2-T tandem solar cells. Indeed, the efficiency increases with improvements in material quality. Additional details on reported efficiencies and stability of 2T TSCs as a function of $E_{gT}$ are available in Fig. ~S10 and Table ~S4 of Supporting Information.\\  

	Current mismatch between the top and bottom sub-cells is a major concern in 2-T solar cells. For the set of parameters, Fig. \ref{eff_estimate}a shows that the $J_{sc}$ first increases and then decreases with the $E_{gT}$. These trends are in accordance with the literature \cite{babics2023temperature} and can be explained as follows: For smaller $E_{gT}$, the performance is limited by the Si bottom cell. As $E_{gT}$ increases the current generated by the Si bottom cell increases, which improves the $J_{sc}$ of the 2-T TSC. At $E_{gT}=1.68\, \mathrm{eV}$, perfect current matching happens between the top and bottom cells. Beyond that, the $J_{sc}$ of the TSC is limited by the top cell and decreases with an increase in $E_{gT}$. The optimal current matching occurs at lower $E_{gT}$ for higher temperature as the perovskite $E_{g}$ increases with temperature (See Fig.~S7 and S8 for more details). Fig. \ref{eff_estimate}b shows that the tandem $V_{oc}$ increases linearly with $E_{gT}$. The  $V_{oc}$ decreases by $\sim 120\, \mathrm{mV}$ for  30 - $40\,^\circ$C increase in the cell temperature. The fill factor (FF) shows an opposing behavior compared to the $J_{sc}$, as observed in Fig. \ref{eff_estimate}c,  and is governed by whether the silicon bottom or perovskite top cell dominates for a given $E_{gT}$. Detailed discussion on the influence of temperature and $E_{gT}$ on tandem cell parameters is provided in the Section 2 of Supporting Information (See Fig~S7 and S8).
       
	The variation of 2-T cell efficiency as a function of $E_{gT}$ is shown in Fig. \ref{eff_estimate}d. A quick comparison with Fig.~\ref{fig:phase}a indicates that the maximum achievable practical efficiency in this configuration is significantly lower than SQ limits. This is mainly due to the non-radiative recombination in both cells and the efficiency could improve with material quality (see Fig. S9 of Supporting Information for details). The simulation results indicate that the optimal $E_{g}$ for the top cell is $\sim 1.68\, \mathrm{eV}$, much different from the SQ prediction of $1.73\, \mathrm{eV}$. In addition, we also find that the 2-T efficiency shows a much lower sensitivity with $E_{gT}$. In the range of $1.58-1.77\, \mathrm{eV}$, the efficiency at 300 K varies only by 5\%. At a higher temperature (330 K), the variation is even lower (3\%).

	The reduced sensitivity of efficiency on $E_{gT}$ can be explained using Fig.~\ref{eff_estimate}a. For the SQ limit calculations, the current mismatch between the top and bottom cells ranges from $8$ to $-2\, \mathrm{mA/cm^2}$. In contrast, for TMM calculations with practical device dimensions and material properties, the current mismatch varies from $4$ to $-4\, \mathrm{mA/cm^2}$ over the same range of $E_{gT}$. This reduction in current mismatch is the key factor that contributes to the reduced sensitivity of efficiency with $E_{gT}$. Additional details regarding the current mismatch and its dependence on $E_{gT}$ and temperature are provided in Section 2 of Supporting Information (Fig.~S5).\\
	
	The results shown in Fig.~\ref{eff_estimate}d indicate that the efficiency decreases with temperature. Interestingly, the decrease in efficiency with temperature is greater for larger $E_{gT}$ (compare the results for 300 K and 330 K in Fig. \ref{eff_estimate}d). As the module temperature is expected to vary significantly under field conditions, this indicates that the EY of 2-T TSCs is expected to show a lower variation with $E_{gT}$ as compared to the corresponding variation in efficiency. This opens opportunities for perovskites with lower $E_{g}$ to be competitive, as detailed in the following sections on the energy yield and LCOE.

\begin{figure*}[ht]
		\centering
		\includegraphics[width=0.95\linewidth]{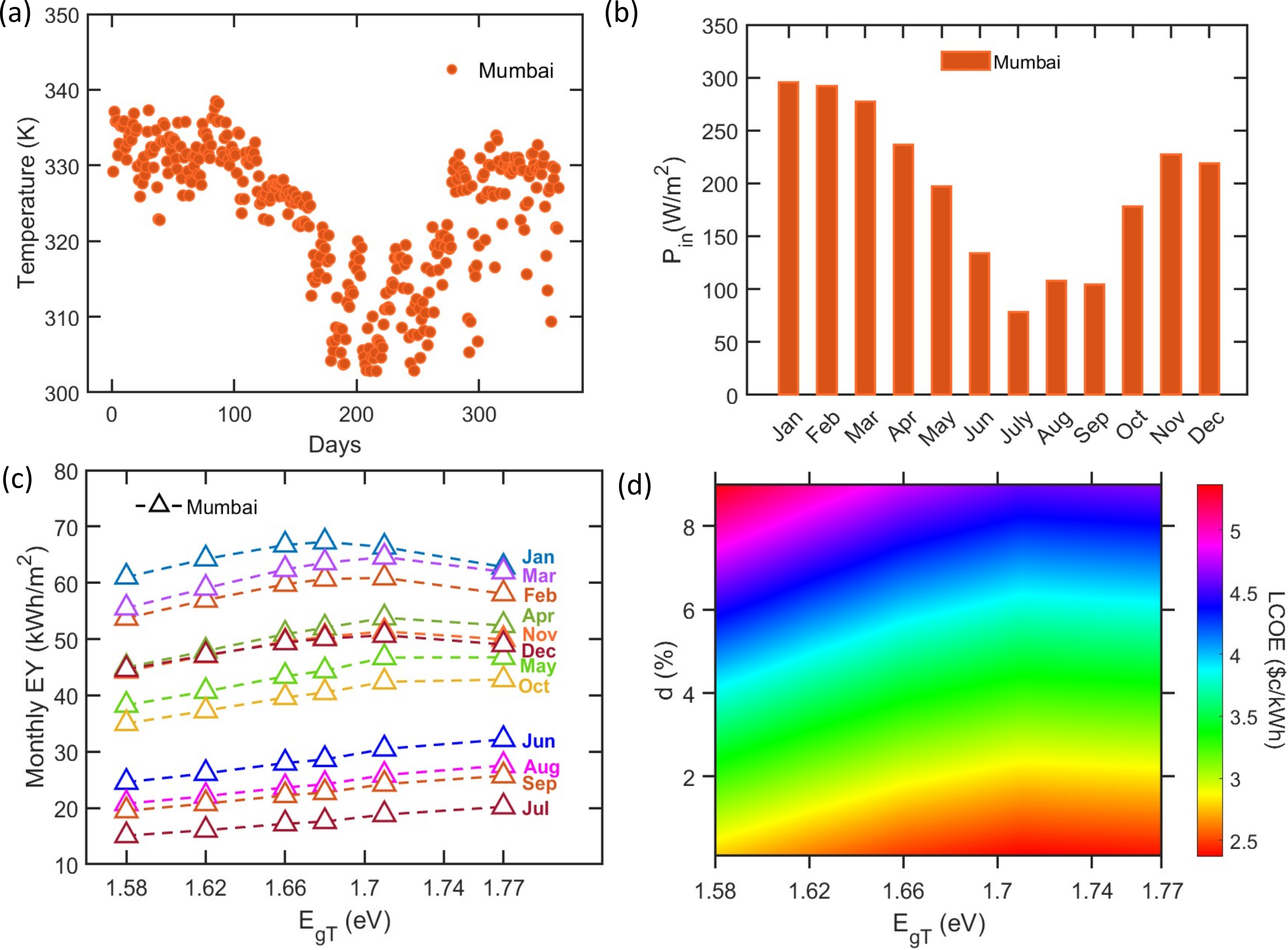}
        \caption{Energy yield and LCOE of  P-Si tandem cells for different \(E_{gT}\) at Mumbai. (a) Calculated maximum module temperature per day and (b) mean incident power density   per month throughout the year (2019). (c) The monthly energy yield estimates for Mumbai. (d) LCOE map as a function of degradation rate and  $E_{gT}$ for 25-year lifespan near Mumbai. }
		\label{fig:3}
	\end{figure*}

\section{Annual Energy Yield}
    
	 The power generated by a solar cell depends on both the temperature and the incident solar spectrum \cite{babics2023temperature}. Consequently, factors such as geographical location and climate conditions influence the cumulative energy output of a PV panel   \cite{horantner2017predicting,jahangir2024planet}. Reasonable estimates for the initial EY, accounting for all these factors, can be obtained as
	\begin{equation}
    EY=\int P_{out}(t)dt\\	   
		\label{eq:EY}
	\end{equation}
	where $t$ is the time and $P_{out}$ is the power output from the PV panel at time $t$.  Note that $P_{out}$ depends on $\Phi_{in}(t)$, the incident spectra  and $T$, the module temperature (both at time $t$). The module temperature in turn depends on various parameters like ambient temperature and wind speed.\\ 
	
 Fig. \ref{fig:3} shows the field conditions and expected EY estimates for Mumbai, India. 
In contrast to studies that calculate energy yield using global horizontal irradiance from Typical Meteorological Year (TMY) data for a location \cite{tomvsivc2023energy,singh2025annual}, we utilize spectral on-demand data from the National Solar Radiation Database (NSRDB) \cite{NSRDB}, which provides hourly spectral irradiance and information on climatic conditions.  This approach allows us to calculate power output for each hour ($P_{out}(t)$), explicitly accounting for spectral variations under real field conditions at different locations. The $P_{out}(t)$ is calculated using an equivalent circuit model calibrated with the drift-diffusion simulations to reduce computation complexity (See Fig.~S11,  Section 3 of Supporting Information for more details ). Fig.~\ref{fig:3}a shows the estimated maximum module temperature for each day of the year in Mumbai (day 0 corresponds to January 1). The module temperature ($T_{m}$) was calculated using King's model, as described in Section 3 of Supporting Information. Fig.~\ref{fig:3}b depicts the average power density of the incident spectrum ($P_{in}$) estimated for each month at Mumbai (See Fig.~S14 and S15 for temperature and $P_{in}$ variation at other locations). The temporal variation in module temperature and $P_{in}$ reflects the climatic condition of the chosen city. Mumbai enjoys good monsoon showers during June-September and the measured data reflects the same - in terms of lower $P_{in}$ and module temperature. \\
     
	Fig. \ref{fig:3}c illustrates the monthly energy yield of P-Si 2-T TSCs at Mumbai for various  $E_{gT}$. For every $E_{gT}$, the efficiency of the 2-T cell was numerically estimated at the corresponding module temperature.	The hourly power output is then estimated using the methodology described in Sec.3 of the Supporting Information. Here, we use the hourly incident power spectrum as the $\Phi_{in}$ for the top cell, TMM calculations to estimate the photo-generation rates, and numerical simulations at the corresponding temperature (determined using King's model) to estimate the module efficiency and hence the power output at each hour of the year. Further, as indicated by eq. \ref{eq:EY}, the monthly energy yield is calculated by integrating the hourly power output over the corresponding month. Consistent with the field conditions reported in Figs. \ref{fig:3}a,b, we find that the monthly EY is at a minimum during the monsoon season in Mumbai (see Fig. \ref{fig:3}c). Further, the P-Si 2T TSCs yield best monthly EY during January-March when the solar insolation is high. In addition, we find that the incident spectrum has a crucial role on the optimal $E_{gT}$.\\
    
    As the energy yield depends on $E_{gT}$, it is important to take into account the degradation rates in the selection criteria for the top cell perovskite. Indeed, while the cost of the tandem cell might not depend much on $E_{gT}$, the degradation rates could vary with $E_{gT}$ - especially given the fact that higher Br content is required in wide $E_{g}$ perovskites \cite{unger2017roadmap}. In fact, LCOE estimates can provide better quantitative insights regarding the influence of degradation rate on the optimal $E_{gT}$. The LCOE, defined as the ratio of the lifetime cost to the lifetime energy yield over the module's lifetime, is 
	\begin{equation}
		LCOE=\frac{C_{f}(1+\sum_{t=0}^{n}X(1+r)^{-t})}{\sum_{t=0}^{n}EY(1-d)^{t}(1+r)^{-t}}
		\label{eq:LCOE}
	\end{equation}
	where $ C_{f}$ represents the fixed cost of module and land \cite{patel2019lcoe}. (See Fig.~S12 for more details). Here, we assume that the time dependent component of costs is a certain percentage (\( X \)) of the initial fixed cost - in tune with recent literature \cite{zafoschnig2020race}. \(X\) is reported to be  $\sim1\%$, though it may depend on geographical location \cite{zafoschnig2020race}. The parameters used in subsequent analysis are well calibrated with recent literature (ref. \cite{zafoschnig2020race,patel2019lcoe}, see Sec 4 of Supporting Information).
    
	Fig.~\ref{fig:3}d shows the variation of LCOE of 2-T P-Si TSCs evaluated at Mumbai as a function of \(d\) and \(E_{gT}\) for a lifespan of 25 years. The LCOE is calculated using the annual EY estimated using the field data (i.e., sum of the monthly EY shown in Fig. \ref{fig:3}c). As expected, the results indicate that lower degradation rates are needed to achieve lower LCOEs. Surprisingly, these results also indicate that the optimal $E_{gT}$ depends on the degradation rate. For $2 < d <3$,  the LCOE values are similar for the \(E_{gT}\) range under consideration. For smaller as well as larger $d$, the LCOE varies significantly across the $E_{gT}$ range.
    
	\section{Geographic location specific optimal $E_{gT}$ and target degradation rates}
\begin{figure}
    \centering
    \includegraphics[width=0.90\linewidth]{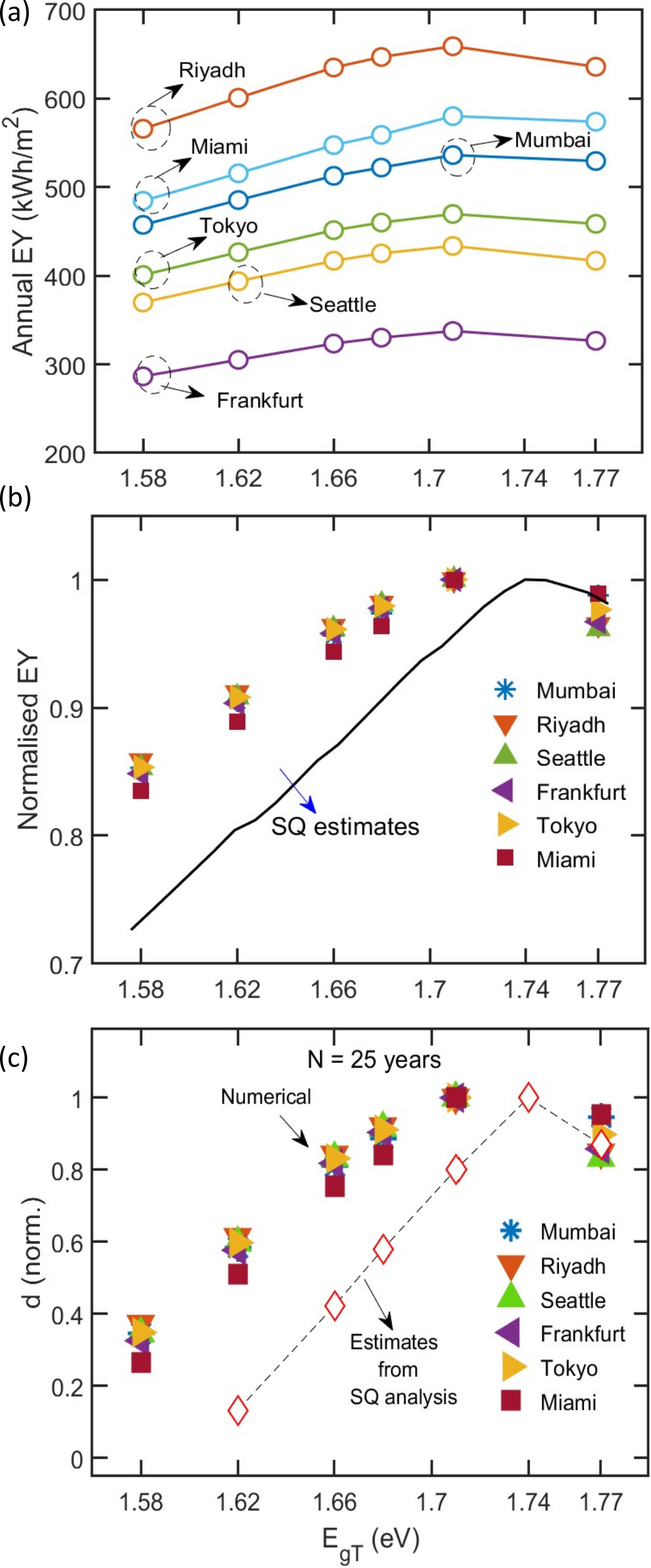}
    \caption{Energy yield and stability targets for P-Si tandem solar cells at different locations. (a) Annual EY vs. $E_{gT}$, (b) Comparison of normalized EY - numerical (symbols) vs. estimates based on SQ analysis (solid line) and (c) Normalized degradation rates ($d_{\text{norm.}}$) which results in similar EY. The numerical simulation results are normalized against $d_1=3\%$ for $E_{gT}=1.71\,\mathrm{eV}$ and $N=25$ years. The SQ estimates are normalized against $d_1=3\%$ for $E_{gT}=1.73\,\mathrm{eV}$ and $N=25$ years. 
    }
    \label{fig:target_rate}
\end{figure}

    Figure \ref{fig:target_rate}a compares the annual EY estimates for different cities across the world. The chosen cities have diverse climatic conditions and hence the energy yield estimates  reflect the range of real-world environments where tandem solar cells are expected to perform.  Here also we used measured data (i.e., fixed-tilt spectral irradiance and climatic conditions) from the National Solar Radiation Database (NSRDB) \cite{NSRDB}. The calculations have been performed for the respective ambient conditions using the same methodology adopted for Mumbai (as detailed in previous section). For every city, we find that the maximum annual EY is achieved for a band gap of $E_{gT} \sim 1.71\,\mathrm{eV}$ - which is different from the predictions of SQ analysis (of $1.73\,\mathrm{eV}$) and efficiency calculations shown in Fig. \ref{eff_estimate}a ($\sim1.68\,\mathrm{eV}$). This difference is clearly due to the variation in the spectrum and ambient conditions. The monthly EY provided in Fig. \ref{fig:3}c offers additional insights in this regard. For months with high incident irradiance, Fig. \ref{fig:3}c indicates that the optimal $E_{gT}$ is $\sim 1.68\, \mathrm{eV}$. For months with the lowest irradiance, the optimal $E_{gT}$ is $\sim 1.77\,\mathrm{eV}$. Consequently, the annual EY shows a peak near $E_{gT} \sim 1.71\,\mathrm{eV}$. \\

    Figure \ref{fig:target_rate}b compares the normalized EY at various cities. Interestingly, despite the enormous variations in climatic conditions and spectral irradiance, the variation of normalized EY with the $E_{gT}$ is nearly the same for all the chosen cities. This simplifies a design challenge for P-Si 2T tandems solar cells - the relative performance of such TSCs is nearly independent of geographic locations. As such, degradation rates could be the key parameter that determines economic viability. \\
    
    Our numerical simulations also indicate that location-specific estimates for annual EY are less sensitive to $E_{g,T}$ - in contrast to the estimates based on SQ analysis (compare between the symbols and solid line in Fig. \ref{fig:target_rate}b). The reduced sensitivity of the numerically estimated EY on $E_{gT}$ is due to the increase in the temperature dependence of efficiency at higher $E_{gT}$ values (see Fig. \ref{eff_estimate}d and Fig.~S8 of Supporting Information). This trend is significantly influenced by the temperature dependence of current mismatch. Interestingly, the temperature dependence increases for $E_{gT}$ in the range $1.68 -1.71\, \mathrm{eV}$, where the current mismatch is negligible at room temperature. However, even minor temperature-induced changes could lead to a significant current mismatch at high temperatures (see Fig.~S5 and S8 of Supporting Information). As a result, the higher temperature dependence partially offsets the efficiency advantage of larger $E_{gT}$, thereby reducing the overall sensitivity of the energy yield to $E_{gT}$. This finding highlights that a wider range of perovskite materials could serve as viable candidates for the top cell. \\
    
Fig.~\ref{fig:target_rate}c provides a thumb rule or selection criterion for the optimal perovskite in terms of normalized degradation rate and $E_{gT}$. Here the Y axis denotes normalized degradation rates at which a P-Si tandem solar cell with $E_{gT}$ gives the same LCOE as a reference 2T P-Si TSC with $E_{gT}=1.71\,\mathrm{eV}$ (i.e, for same N). Note that the results in Fig. \ref{fig:target_rate}a,b indicate that EY is maximum near $E_{gT} \sim 1.71\, \mathrm{eV}$. Hence, to achieve the a specific LCOE for same N, the degradation rate for other $E_{gT}$ has to be smaller than its value at $E_{gT}=1.71\,\mathrm{eV}$ (as predicted by eq. \ref{eq:LCOE}). Given that the normalized EY estimates show near similar variations (see Fig. \ref{fig:target_rate}b), we find that the target degradation rates are also similar across the chosen cities.  A comparison with the corresponding SQ limits quantify the influence of local incident spectra and temperature on target degradation rates. For example, with $N=25$ years, under field conditions, a TSC with $E_{gT,1}=1.62\, \mathrm{eV}$ can yield the same LCOE as a TSC with $E_{gT,2}=1.71\, \mathrm{eV}$, when $d_1=0.6d_2$ (see solid symbols in Fig.~\ref{fig:target_rate}c). On the other hand, EY estimates based on SQ analysis demand $d_1=0.3d_2$  to achieve the same LCOE, which presents a more stringent demand. These estimates are also influenced by the value of $N$, the lifetime. Figure S13 of Supporting information describes similar criteria for degradation rates with $N = 18$ years. Further, any relative improvement in recombination parameters could influence this selection criteria on degradation rates (see Fig.~S9 of Supporting Information).
	\section{Discussions}
 
	Several insights are shared in this manuscript. First, through detailed numerical simulations and analysis, we showed that perovskites with a broad range of $E_{gT}$ are suitable as the top cell for TSCs. This is in stark contrast to the traditional SQ based analysis which identifies $E_{gT}\sim 1.73\, \mathrm{eV}$ as the optimal material. Further, the SQ analysis predicts that 2-T TSC efficiency decreases by as much as 7\% per $0.1\, \mathrm{eV}$ decrease of $E_{gT}$ from its predicted optimal value. Significantly, these are overestimates and underplay the potential of perovskites with lower $E_{gT}$. On the other hand, our analysis relies on experimentally observed temperature dependence of various parameters such as refractive indices and $E_{gT}$. In addition, we account for various non-linearities in carrier transport, recombination, and ion migration in our numerical simulations to arrive at better estimates for the influence of $E_{gT}$ on 2-T TSC efficiency. The EY trends obtained for various cities rely on the measured field data in terms of spectra, temperature, wind speed, etc. Hence availability of better quality data can indeed improve the numerical calculations and subsequent estimates.  \\
    
    Our analysis provides more quantitative information on the notion of optimal $E_{gT}$ for P-Si tandem solar cells. The traditional SQ analysis indicates that $1.73 \, \mathrm{eV}$ is the ideal $E_{gT}$. However, efficiency calculations with AM1.5G as the incident spectrum and temperature dependence for various parameters like refractive indices suggest that the optimal $E_{gT}$  $\sim 1.68\, \mathrm{eV}$ (see Fig. \ref{eff_estimate}). Further, our calculations using field data and explicit temperature dependence for sub-cell characteristics indicate that EY peaks at $E_{gT} \sim 1.71\, \mathrm{eV}$ (see Fig.~\ref{fig:target_rate}a). 
    Given the influence of multiple parameters,  $1.68 -1.72\, \mathrm{eV}$ represents an optimal range of $E_{gT}$ for P-Si 2-T TSCs.\\
	
    Our analysis can be extended to explore the trade-offs involved in the design of 4T tandem solar cells. For a tandem configuration with a silicon bottom cell, SQ analysis shows that the 4-T architecture exhibits only a minimal $\sim2\%$ efficiency variation across the $E_{gT}$ range of 1.5 to 1.8 eV, while the 2-T tandem shows a much larger $\sim15\%$ variation due to current mismatch constraints (see Fig.~S1d). This makes bandgap selection in 4-T tandems much more flexible, allowing the choice to be primarily guided by the stability of the top cell material. Meanwhile, bifacial 2-T tandems with one-axis tracking can reduce current mismatch and offer an increase in energy yield \cite{khan2021review}. This improvement depends on factors such as ground albedo, geographic location, and device architecture, including the thickness and bandgap of the top cell \cite{de2022bifacial}. It has been reported that bifacial P–Si 2-T tandems with narrow bandgap perovskite top cells show significant gain in power output due to increased absorption in the silicon bottom cell, provided there is sufficient albedo. This reduces current mismatch compared to wider bandgap counterparts and such designs could be more stable due to their low bromine content in perovskites \cite{de2021efficient}.\\
    
  Benchmarking EY predictions against outdoor test data of P–Si tandem solar cells is essential for accurate LCOE estimates\cite{liu202128,pathways,babics2023one}. In practice, these devices often exhibit non-uniform degradation over time, including an early burn-in phase with rapid initial performance loss followed by a slower, more linear decline\cite{domanski2018systematic,zhang2022big}. In contrast, typical LCOE estimates rely on the assumption of linear degradation rates. The insights shared in this manuscript regarding degradation rates are broadly applicable for scenarios in which early phase burn-ins are comparable for different $E_{gT}$. Our methodology can be appropriately adapted to account for non-linear degradation rates. Indeed, calibrated outdoor tests and accelerated aging tests to quantify degradation phenomena are of immense relevance in this regard.\\

Among the wide variety of inorganic--organic metal halide perovskites available for solar cell and other optoelectronic applications, triple-cation mixed-halide perovskites have emerged as the most promising materials and are currently used in state-of-the-art, record-efficiency P--Si tandems \cite{liu2024perovskite, ugur2024enhanced}. These materials offer improved stability by effectively suppressing phase segregation and enhancing structural robustness under thermal and environmental stress conditions \cite{pei2025stability,yang2025reductive,mei2025recent,mosquera2025multifaceted}. Specifically, (Cs$_{0.05}$(MA$_y$FA$_x$)$_{0.95}$Pb(I$_x$Br$_y$)$_3$) enables facile bandgap tunability from 1.5~eV to 1.9~eV by adjusting the bromide (Br) content \cite{peña2020halide}. Such compositional tuning simplifies fabrication and suggests comparable costs for tandem cell designs, though we acknowledge that stability concerns, encapsulation, and balance-of-system costs may depend on $E_{gT}$. This necessitates comparisons based on LCOE (i.e., not based on EY alone) and could be topics of future research.  While we observe that climate-specific design could identify the optimal perovskite for top-cell based on geographic location and relative degradation rates, we note that its implementation can be influenced by practical industry challenges \cite{adothu2024comprehensive}. Diverse product lines and complex supply chains could increase costs and operational complexity. These challenges can be mitigated through modular manufacturing approaches and improved supply chain coordination \cite{yang2024achievements}. Such practices can help the industry effectively balance performance benefits and manufacturing feasibility, thus accelerating the commercialization of high-efficiency tandem solar cells.
    \section{Conclusions}
	In this manuscript, we report annual energy yield estimates for P-Si 2-T TSCs for  perovskite materials with $1.58 \, \mathrm{eV} < E_{gT} < 1.77\, \mathrm{eV}$ at various geographic locations. Our methodology explicitly accounts for the temperature dependence of material and transport parameters as well as the spectral variations under field conditions. We show that the energy yield shows a markedly reduced dependence on $E_{gT}$ as compared to SQ limits. Further, we report that annual EY shows a near-identical scaling with $E_{gT}$ for the cities chosen in this study. In addition, we find a twofold relaxation in the required criteria for degradation rates, as compared to the traditional SQ based analysis, at which low $E_{gT}$ perovskites could be competitive. This indicates that a wide variety of perovskites could yield comparable LCOEs with moderate improvements in degradation rate.
    
    \section{Acknowledgement}
	The authors acknowledge National Center for Photovoltaics Research and Education (NCPRE), IIT Bombay.
	
    \section{Supporting Information}
	Detailed discussions on SQ limit analysis, Numerical simulations, Energy yield and LCOE calculations are provided in Supporting Information.	

	\section*{References}

\bibliography{Ref}
\end{document}